\documentclass[aps,preprint,prd,amsmath,superscriptaddress,amssymb,amsfonts]{revtex4-2}
\usepackage{mathtools}
\usepackage{graphicx}
\usepackage{subfigure}%
\usepackage{float}
\usepackage{booktabs}
\usepackage{mathrsfs}
\usepackage{tensor}
\usepackage{verbatim}
\usepackage[colorlinks=true,
linkcolor=blue, %%%修改此处为你想要的颜色
anchorcolor=blue, %%修改此处为你想要的颜色
citecolor=blue, %%设置文中引用参考文献的颜色
urlcolor=blue%%设置DOI的颜色为蓝色
]{hyperref} 

\newcommand{\vnabla}{\overset{v}{\nabla}}
\date{\today}
\begin{document}
	\title{Effective ray equations for vortex light and their application in an optical waveguide}
	\author{Wei-Si Qiu}
	\email{qiuws@mail2.sysu.edu.cn}
	\affiliation{School of Physics and Astronomy, Sun Yat-sen University, 519082 Zhuhai, China}
	\author{Dan-Dan Lian}
	\email{liandd@mail.sysu.edu.cn}
	\affiliation{School of Physics and Astronomy, Sun Yat-sen University, 519082 Zhuhai, China}
	\author{Peng-Ming Zhang}
	\email{zhangpm5@mail.sysu.edu.cn}
	\affiliation{School of Physics and Astronomy, Sun Yat-sen University, 519082 Zhuhai, China}

\begin{abstract}

Beyond its spin, light can also carry intrinsic orbital angular momentum (IOAM), termed as vortex light. In this study, we derive effective ray equations for vortex light by applying the WKB approximation to the covariant Maxwell equations. According to these equations, the propagation of vortex light can be significantly affected by its IOAM, as suggested by numerous studies. To examine the effects of IOAM, we solve the effective ray equations for vortex light and investigate its ray trajectory within a specific optical waveguide. Our findings indicate that the ray trajectory of vortex light exhibits a divergence perpendicular to the normal propagation plane, akin to the spin Hall effect in light. This divergence, termed as the orbital Hall effect, stems from the IOAM of the light. In this study, the effective ray equations are derived by modeling the interaction between light and media as light's free fall in a curved spacetime. Therefore, observing the orbital Hall effect could not only  enhance our understanding of light's spin and IOAM, but also offer novel insights into the coupling between light and gravitational fields.

\end{abstract}

\maketitle
\newpage

\section{Introduction}\label{introduction}
%相比之前区分了Hamilton's equation, Hamilton-Jacobi equation,Hamilton-Jacobi system, ray equations,ray trajectories的术语%
%补充了很多关于领头阶的内容，是结合上次退稿的反馈%
%把原来appendix A的部分挪到主体段，考虑到IOAM-dependent的推导是本文主要内容之一%
%加了文献%
%IOAM-EOAM 术语用在什么场合合适%
%hamiton‘s equations 的结果讨论段落逻辑顺序更改%
%区分deviate,separate,deflect,split%
%调整章节、标题,部分挪动%
%有很多地方只有公式，但是不知道怎么扩展开来写文字%

The Spin Hall Effect (SHE) of light, resulting from spin-orbit coupling in specific media or external fields, leads to spin-dependent ray trajectories, well-documented in both theory and experiments. Bliokh's introduction of the geometric phase clarifies the light's spin-dependent effects in inhomogeneous isotropic media \cite{bliokh2009spin,Bliokh_2006}. Additionally, Mieling et al. explored polarized light's propagation in curved optical fibers, extending applications in optical waveguides \cite{Mieling_2023}. Extensive experimental work further discusses the SHE of light, demonstrating its detection and observational capabilities \cite{fu2019spin,sheng2023photonic,ling2017recent,hosten2008observation,bliokh2008geometrodynamics,Bliokh_2015,PhysRevA.84.043806,PhysRevLett.102.123903}.

Light can manifest in vortex states, each carrying an Intrinsic Orbital Angular Momentum (IOAM) of $\ell \hbar$ per photon, aligned with its average momentum. These states, known as "vortex" light, are characterized by a wave function $\psi(\mathbf{r}) \propto \exp(i \ell \phi) $, where $\phi$ denotes the azimuthal angle \cite{bliokh2015spin}. Allen et al. first identified vortex light in free-space in 1992 \cite{beijersbergen1993astigmatic}. Analogous to spin, IOAM can also affect light's propagation in media and induce a divergence perpendicular to the plane of propagation defined by the absence of angular momentum \cite{bliokh2017theory}. This phenomenon, known as the Orbital Hall Effect (OHE), has been studied for vortex light along spatial curves \cite{PhysRevA.97.033843} and for free-falling vortex particles in curved spacetime \cite{lian2024motion,qiu2024orbital}.

It has been pointed out that the interaction between light and optical media can be modeled as light's free fall in curved spacetime \cite{BIALYNICKIBIRULA1996245,narimanov2009optical,genov2009mimicking,cheng2010omnidirectional,yang2012electromagnetic,sheng2013trapping,PhysRevD.74.021701}. This suggests the feasibility of using unified effective equations to describe light propagation in both optical media and gravitational fields within the framework of general relativity. The dynamics of particles with angular momentum are typically governed by the Mathisson-Papapetrou-Dixon (MPD) equations \cite{Mathisson1937,papapetrou1951spinning,dixon1970dynamics}. The SHE of light in curved spacetimes has been extensively studied \cite{corinaldesi1951spinning,souriau1974modele,saturnini1976modele,duval2019gravitational}. Ramírez and Deriglazov derived a Hamiltonian with a spin-orbit interaction term, enhancing our understanding of this effect \cite{Ram_rez_2017,Deriglazov_2018}. Furthermore, the SHE in curved spacetime has been explored for exotic photons, which are massless, chargeless particles with anyonic spin, magnetic moments, and exotic charges related to the Carroll group \cite{PhysRevD.106.L121503}.

The IOAM of light is characterized by a space-dependent phase factor $\exp(i\ell \phi)$ in its wave function, while its spin is represented by a polarization vector $\mathbf{\varepsilon}$. This distinction suggests that the MPD equations might not fully capture the effects of IOAM, as they overlook the crucial differences between spin and IOAM. To address this, new effective equations are necessary. This work aims to derive effective ray equations for vortex light from the covariant Maxwell equations. As an application, we investigate the propagation of vortex light in an optical waveguide, detailing the effect of IOAM, known as the OHE. Since the interaction between light and optical media has been modeled as light's free fall in curved spacetime, observing the OHE in optical waveguides might also provide new insights into the coupling of light with gravity.

To achive our purpose, we utilize the Wentzel-Kramers-Brillouin (WKB) approximation on the covariant Maxwell's equations. This technique was initially introduced by Frolov et al. \cite{PhysRevD.84.044026,PhysRevD.86.024010} and further refined by Yoo \cite{yoo2012notes} and Dolan et al. \cite{dolan2018geometrical,dolan2018higher}. Applying the WKB approximation has allowed for the investigation of the SHE of light in curved spacetimes, as explored by Oancea et al. \cite{oancea2020gravitational,Dahal:2022gop,Andersson:2023bvw,Oancea:2023ylb}. Furthermore, this approximation has also been applied to study the propogation of gravitational waves \cite{Kubota:2023dlz}. 

The structure of this paper is organized as follows: Section \ref{WKB approximations} derives the effective ray equations for vortex light, applying the WKB approximation to the covariant Maxwell equations. This approach effectively describes vortex light propagation in optical media. In Section \ref{material candidates}, we investigate the ray trajectory of vortex light propagating through a specific optical waveguide. Our findings show that its IOAM can induce a divergence perpendicular to the propagation plane, a phenomenon commonly known as the OHE. The paper concludes with a discussion in Section \ref{conclusions}, summarizing the findings and their implications for future research.

In this work, we use the metric signature $(-,+,+,+)$ and adopt relativistic units where $c=\hbar=G=1$, with $c$ representing the speed of light. The phase space is defined as the cotangent bundle $T^*M$, with points in phase space represented by $(x,p)$. Greek indices span the four coordinates in a general coordinate system, while Latin indices $i, j, k, . . .$ are confined to the three spatial coordinates. The Riemann curvature tensor is given by $R^\alpha_{\ \mu \nu \beta}=\partial_{\beta} \Gamma_{\mu \nu}^{\alpha}-\partial_{\nu} \Gamma_{\mu \beta}^{\alpha}+\Gamma_{\mu \nu}^{\gamma} \Gamma_{\beta \gamma}^{\alpha}-\Gamma_{\mu \beta}^{\gamma} \Gamma_{\nu \gamma}^{\alpha}$, where $\Gamma_{\mu \nu}^{\alpha}$ denotes the Christoffel symbols. Additionally, we employ the $\mathcal{O}$ notation to indicate that a scalar function $f$, dependent on a parameter $\epsilon$, satisfies  $f(\epsilon) = \mathcal{O}(\epsilon^\alpha)$ if for small $\epsilon$, there exists a constant $N$ such that $|f(\epsilon)| \leq N \epsilon^{\alpha}$.

\section{Effective ray equations for vortex light}\label{WKB approximations}

The metric of a curved spacetime can be given by its line element $\text{d}s^2=-g_{\mu\nu}\text{d}x^\mu\text{d}x^\nu$.  And the dynamics of light in such curved spacetime are governed by the covariant Maxwell's equations, expressed as:
\begin{equation} \label{eq:Maxwell_F}
	\nabla{}^\alpha\mathcal{F}{}_{\alpha \beta} = 0,
\end{equation}
where $\mathcal{F}_{\alpha \beta}=\nabla_{\alpha}\mathcal{A}_\beta-\nabla_{\beta}\mathcal{A}_\alpha $ represents the electromagnetic field tensor, and $\mathcal{A}_\alpha$ is the electromagnetic four-potential. Simplifying Maxwell's equations yields:
\begin{equation}\label{eq:Maxwell_F2}
	\hat{D}_{\alpha}{ }^{\beta} \mathcal{A}_{\beta}=0, \quad \hat{D}_{\alpha}{ }^{\beta}=\nabla^{\beta} \nabla_{\alpha}-\delta_{\alpha}^{\beta} \nabla^{\mu} \nabla_{\mu},
\end{equation}
Adopting the covariant Lorenz gauge, $\nabla_\alpha \mathcal{A}^\alpha  = 0$, enables the derivation of Maxwell's equations as Euler-Lagrange equations from the action \cite{oancea2020gravitational}:
\begin{equation} \label{eq:Maxwell_action}
	\begin{split}
		J = \frac{1}{4} \int_M \sqrt{-g}\mathcal{F}_{\alpha \beta} \mathcal{F}^{\alpha \beta}\mathrm{d}^4 x  = \frac{1}{2} \int_M  \sqrt{-g} \mathcal{A}^\alpha \hat{D}\indices{_\alpha^\beta} \mathcal{A}_\beta \mathrm{d}^4 x= \int_M \sqrt{-g} \mathcal{L} \mathrm{d}^4 x,
	\end{split}
\end{equation}
where $\mathcal{L}=\frac{1}{2} \mathcal{A}^\alpha \hat{D}\indices{_\alpha^\beta} \mathcal{A}_\beta$ represents the Lagrangian for electromagnetic fields, and $g=\text{det}(g_{\mu\nu})$ is the determinant of the metric tensor $g_{\mu\nu}$.

Typically, the phase variation of light occurs on a much shorter timescale compared to changes in the metric of curved spacetime. This discrepancy allows us to apply the WKB approximation to the electromagnetic four-potential $\mathcal{A}_\alpha$, formulated as:
\begin{align} \label{eq:WKB_Maxwell}
		&\mathcal{A}_\alpha (x) = \mathrm{Re} \left[ A_\alpha(x, k(x), \epsilon)e^{i\ell \phi} e^{i S(x) / \epsilon} \right],\nonumber\\
		& A_\alpha(x, k(x), \epsilon) = {A_0}_\alpha(x, k(x)) + \epsilon {A_1}_\alpha(x, k(x)) + \mathcal{O}(\epsilon^2),
\end{align}
where $\epsilon$ represents a small expansion parameter, $k_\mu (x) = \nabla_\mu S(x)$ denotes the phase gradient, and $S(x)$ is a real scalar function that depends on $x^\mu$. The complex amplitude $A_\alpha(x, k(x), \epsilon)$ and the phase function $S(x)$  vary slowly relative to the rapid oscillations of the phase factor $e^{i S(x) / \epsilon}$. Notably, the IOAM of vortex light has been characterized by the phase term $e^{i\ell \phi}$, where $\ell$ takes integer values $0, \pm1, \pm2, \pm3, \cdots$.

Given that $\epsilon \ll 1$, the phase of the vector potential $\mathcal{A}_\alpha(x)$ oscillates much more rapidly compared to the amplitude $A_\alpha(x, k(x), \epsilon)$. Consequently, the frequency of light as perceived by an observer is approximately determined by:
\begin{equation}\label{eq-para}
    \omega = -\frac{t^\alpha k_\alpha}{\epsilon},
\end{equation}
where $t^\alpha =dy^\alpha/d\tau$ represents the velocity vector field of the observer. This relationship suggests that the expansion parameter $\epsilon$ is not an arbitrary parameter, but rather, it is directly related to the observed frequency of light.

Under the WKB approximation as specified by Eq. \eqref{eq:WKB_Maxwell}, the Lagrangian $\mathcal{L}$ is expressed as
\begin{equation}
	\mathcal{L}=\mathcal{L}(S,\nabla_\mu S, A_\alpha,\nabla_\mu A_\alpha,A^{*\alpha},\nabla_\mu A^{*\alpha}).
\end{equation}
In this formulation, the real scalar function $S(x)$ is considered an independent field within the Lagrangian $\mathcal{L}$. Consequently, the complex amplitude $A_\alpha$ cannot be regarded as independent from $A^{*\alpha}$, as both depend on $\nabla_\mu S(x)$. The Euler-Lagrange equations derived from $\mathcal{L}$, as detailed by Oancea et al. \cite{oancea2020gravitational}, are given by:
\begin{align}
	\frac{\partial \mathcal{L}}{\partial A^{* \alpha}}-\nabla_{\mu} \frac{\partial \mathcal{L}}{\partial \nabla_{\mu} A^{* \alpha}} & =\mathcal{O}(\epsilon^{2}), \\
		\frac{\partial \mathcal{L}}{\partial A_{\alpha}}-\nabla_{\mu} \frac{\partial \mathcal{L}}{\partial \nabla_{\mu} A_{\alpha}} & =\mathcal{O}(\epsilon^{2}), \\
		\frac{\partial \mathcal{L}}{\partial S}-\nabla_{\mu} \frac{\partial \mathcal{L}}{\partial \nabla_{\mu} S} & =\mathcal{O}(\epsilon^{2}).
\end{align}

Considering Eq. \eqref{eq:WKB_Maxwell}, the Euler-Lagrange equations simplify to
\begin{align}
	& D\indices{_\alpha^\beta} A_\beta - i \epsilon (\nabla_\mu A_\beta+A_\beta e^{-i \ell \phi} \nabla_{\mu}e^{i \ell \phi})\vnabla{}^\mu D\indices{_\alpha^\beta}  - \frac{i \epsilon}{2} A_\beta  \nabla_\mu \vnabla{}^\mu D\indices{_\alpha^\beta} = \mathcal{O}(\epsilon^{2}), \label{eq:HJ1_fullA} \\
	&D\indices{_\alpha^\beta} A^{*\alpha} + i \epsilon \left(\vnabla{}^\mu D\indices{_\alpha^\beta} \right) (\nabla_\mu A^{*\alpha}-A^{*\alpha}e^{-i \ell \phi} \nabla_{\mu}e^{i \ell \phi}) + \frac{i \epsilon}{2} \left( \nabla_\mu \vnabla{}^\mu D\indices{_\alpha^\beta} \right) A^{*\alpha} =\mathcal{O}(\epsilon^{2}), \label{eq:HJ2_fullA} \\
	&\nabla_\mu \left\{ A^{*\alpha} A_\beta \vnabla{}^\mu D\indices{_\alpha^\beta}   
	- \frac{i\epsilon}{2} ( A^{*\alpha} \nabla_\nu A_\beta - A_\beta \nabla_\nu A^{*\alpha}+2A_\beta A^{*\alpha}e^{-i \ell \phi} \nabla_{\nu}e^{i \ell \phi})\vnabla{}^\mu \vnabla{}^\nu D\indices{_\alpha^\beta}\right\} = \mathcal{O}(\epsilon^{2}) \label{eq:transp_fullA},
\end{align}
where 
\begin{equation}
	D\indices{_\alpha^\beta}=\frac{1}{2} k_{\mu} k^{\mu} \delta_{\alpha}^{\beta}-\frac{1}{2} k_{\alpha} k^{\beta},\quad \vnabla{}^\mu D_{\alpha}^{\ \beta}=k^{\mu} \delta_{\alpha}^{\beta}-\frac{1}{2} \delta_{\alpha}^{\mu} k^{\beta}-\frac{1}{2} g^{\mu \beta} k_{\alpha}.
\end{equation}
Here $D\indices{_\alpha^\beta}$ represents the symbol of the operator $\hat{D}\indices{_\alpha^\beta}$, and $\vnabla{}^\mu D\indices{_\alpha^\beta}$ denotes the vertical derivative of $D\indices{_\alpha^\beta}$, both evaluated at the phase space point $(x,p)=(x,k)$. Correspondingly, the covariant Lorenz gauge condition simplifies to
\begin{equation}\label{eq:Lorenz_cond1}
 k_\alpha {A_0}^{\alpha}   = 0,\quad
		\nabla_\alpha {A_0}^\alpha + i k_\alpha {A_1}^\alpha+{A_0}^\alpha \cdot e^{-i \ell \phi} \nabla_{\alpha}e^{i \ell \phi}  = 0,
\end{equation}
when considering terms up to first order in $\epsilon$. 

Neglecting terms of order higher than the zeroth order in $\epsilon$, the Euler-Lagrange equations reduce to
\begin{equation}\label{transport equation1}
		D_{\alpha}^{\beta} A_{0 \beta}=0,\quad D_{\alpha}^{\beta} A_{0}^{* \alpha}=0,\quad \nabla_{\mu}\left[\left(\nabla^{\mu} D_{\alpha}^{\beta}\right) A_{0}^{* \alpha} A_{0 \beta}\right]=0.
\end{equation}
Assuming  $k_a$ is a null covector, we derive
\begin{align}\label{transport equation2}
			k_\mu k^\mu=0,\quad k^{\alpha} A_{0 \alpha}=k_{\alpha} A_{0}^{* \alpha}=0,\quad\nabla_{\mu}\left(k^{\mu} A_{0 \alpha}A_{0}^{*\alpha} \right)=0.
\end{align}
As detailed by Oancea et al. \cite{oancea2020gravitational}, these equations lead to the effective ray equations for light in geometrical optics. Our aim is to investigate the influence of IOAM on the propagation of vortex light, necessitating the derivation of effective ray equations beyond the zeroth-order geometrical optics.

\subsection{First-order geometrical optics}

Up to first order in $\epsilon$, the Euler-Lagrange equations \eqref{eq:HJ1_fullA}, \eqref{eq:HJ2_fullA}, and \eqref{eq:transp_fullA} simplify as follows:
\begin{align}
		&D\indices{_\alpha^\beta} {A_1}_\beta - i \left(\vnabla{}^\mu D\indices{_\alpha^\beta} \right) (\nabla_\mu {A_0}_\beta+{A_0}_\beta e^{-i \ell \phi} \nabla_{\mu}e^{i \ell \phi}) 
		- \frac{i}{2} \left( \nabla_\mu \vnabla{}^\mu D\indices{_\alpha^\beta} \right) {A_0}_\beta = 0, \label{eq:transp_a0}\\
		&D\indices{_\alpha^\beta} {A_1}^{*\alpha} + i \left(\vnabla{}^\mu D\indices{_\alpha^\beta} \right) (\nabla_\mu {A_0}^{*\alpha}-{A_0}^{*\alpha}e^{-i \ell \phi} \nabla_{\mu}e^{i \ell \phi}) 
		+ \frac{i}{2} \left( \nabla_\mu \vnabla{}^\mu D\indices{_\alpha^\beta} \right) {A_0}^{*\alpha} = 0,\label{eq:transp_a0*}\\
		&\nabla_{\mu}\left[
		\frac{1}{2} g^{\mu \nu}\left(A_{0}^{* \alpha} \nabla_{\nu} A_{0 \alpha}-A_{0 \alpha} \nabla_{\nu} A_{0}^{* \alpha}\right)+ A_{0}^{* \alpha} A_{0 \alpha} e^{-i \ell \phi} \nabla^{\mu}e^{i \ell \phi}  \right]=0.\label{transport equation3}
\end{align}
By contracting Eq.~\eqref{eq:transp_a0} with $A^{*\alpha}$ and Eq.~\eqref{eq:transp_a0*} with $A_\beta$, and then combining them, we obtain
\begin{align} \label{eq:efective_disp_2}
		&\quad\frac{1}{2} k_\mu k^\mu ( {A_0}^{*\alpha} {A_0}_\alpha + \epsilon {A_0}^{*\alpha} {A_1}_\alpha + \epsilon {A_1}^{*\alpha} {A_0}_\alpha ) 
		- \frac{i \epsilon}{2} k^\mu \left( {A_0}^{*\alpha} \nabla_\mu {A_0}_\alpha - {A_0}_\alpha \nabla_\mu {A_0}^{*\alpha} \right) \nonumber\\
		&-i\epsilon |{A_0}|^2 k^\mu e^{-i \ell \phi} \nabla_{\mu}e^{i \ell \phi} = 0,
\end{align}
accounting for terms up to first order in $\epsilon$. 

The intensity of light is typically represented by $\mathcal{I} = \mathcal{A}^{*\alpha} \mathcal{A}_\alpha$. Considering terms up to the first order in $\epsilon$, the intensity $\mathcal{I}$ can be expressed as
\begin{equation} \label{eq:def_I}
		\mathcal{I} \simeq {A_0}^{*\alpha} {A_0}_\alpha + \epsilon ({A_0}^{*\alpha} {A_1}_\alpha + {A_1}^{*\alpha} {A_0}_\alpha).
\end{equation}
Accordingly, the complex amplitude $A_\alpha$ can be described using the intensity $\mathcal{I}$ and approximated by
\begin{equation}\label{amplitude}
	A_\alpha = \sqrt{\mathcal{I}}a_\alpha\simeq \sqrt{\mathcal{I}}\left( {a_0}_\alpha + \epsilon {a_1}_\alpha \right),
\end{equation}
where $a_\alpha$ denotes a unit polarization complex covector that fulfills the condition ${a}^{*\alpha} {a}_\alpha = 1$.

Taking into account Eqs.~\eqref{amplitude} and \eqref{eq:def_I}, we can simplify Eqs. \eqref{eq:efective_disp_2} and \eqref{transport equation3} as follows:
\begin{align}
&\frac{1}{2} g^{\mu \nu} k_{\mu} k_{\nu} -i\epsilon k^{\mu} a_{0}^{* \alpha} \nabla_{\mu} a_{0 \alpha}-i\epsilon k^{\mu} e^{-i \ell \phi} \nabla_{\mu}e^{i \ell \phi} = 0,\label{eq-rela}\\
&\nabla_{\mu}\left\{\mathcal{I}\left[k^{\mu}-i\epsilon g^{\mu \nu}a_{0}^{* \alpha} \nabla_{\nu} a_{0 \alpha}-i\epsilon g^{\mu \nu}e^{-i\ell \phi}\nabla_\nu e^{i\ell \phi}   \right]\right\} =0.\label{eq-trans}
\end{align}
Compared to Eq. \eqref{transport equation2} from geometrical optics, the effective dispersion relation \eqref{eq-rela} and the transport equation \eqref{eq-trans} are influenced by the spin angular momentum (spin) and intrinsic orbital angular momentum (IOAM) of vortex light, as indicated by the terms $a_{0}^{* \alpha} \nabla_{\nu} a_{0 \alpha}$ and $e^{-i\ell \phi}\nabla_\nu e^{i\ell \phi}$, respectively. Thus, it is evident that the propagation of vortex light can be significantly affected by its IOAM.

To examine the effects of spin and IOAM on light propagation, we introduce a unit complex polarization vector, $a'_\alpha$, defined by $a'_\alpha\equiv a_{0\alpha}e^{i\ell\phi}$. This leads to the relation $a'^{*\alpha}\nabla_\mu a'_{\alpha}=  a_{0}^{* \alpha} \nabla_{\mu} a_{0 \alpha}+ e^{-i \ell \phi} \nabla_{\mu}e^{i \ell \phi}$. As a result, the effective dispersion relation \eqref{eq-rela} and the transport equation \eqref{eq-trans} simplify to:
\begin{align}
	&\frac{1}{2} g^{\mu \nu} k_{\mu} k_{\nu} -i\epsilon k^{\mu} a'^{*\alpha}\nabla_\mu a'_{\alpha} = 0,\label{seq-rela}\\
	&\nabla_{\mu}\left\{\mathcal{I}(k^{\mu}-i\epsilon g^{\mu \nu}a'^{*\alpha}\nabla_\mu a'_{\alpha})\right\} =0.\label{seq-trans}
\end{align}

Using the newly defined polarization vector $a'_\alpha$, the effective dispersion relation and transport equation for vortex light each align in form with those for spin-polarized light. Therefore, these equations for vortex light can be decomposed using the method described by Oancea et al. \cite{oancea2020gravitational}. This involves the introduction of the null tetrad $\left\{k_{\alpha}, n_{\alpha}, m_{\alpha}, \bar{m}_{\alpha}\right\}$, characterized by:
\begin{equation}
	\begin{split}
		m_\alpha \bar{m}^\alpha = 1, \qquad k_\alpha n^\alpha = -1, \\
		k_\alpha k^\alpha = n_\alpha n^\alpha = m_\alpha m^\alpha = \bar{m}_\alpha \bar{m}^\alpha &= 0, \\
		k_\alpha m^\alpha = k_\alpha \bar{m}^\alpha = n_\alpha m^\alpha = n_\alpha \bar{m}^\alpha &= 0,
	\end{split}
\end{equation}
enabling the decomposition of $a'_{\alpha}$ as:
\begin{equation}
	a'_\alpha(x, k) = z_1(x) m_\alpha(x, k) + z_2(x) \bar{m}_\alpha(x, k) + z_3(x) k_\alpha(x),
\end{equation}
where  $z_1$, $z_2$, and $z_3$ are complex scalar functions representing the projection coefficients.

Following the analysis presented in Appendix \ref{appendix2}, we can succinctly reformulate the effective dispersion relation \eqref{eq-rela} and the transport equation \eqref{eq-trans} as:
\begin{align}
		&\frac{1}{2} g^{\mu \nu} k_\mu k_\nu +2\epsilon |z_1(0)|^2 \partial_{\mu}\gamma -\epsilon (\sigma+\ell) k^\mu B_\mu=0,\label{eq-rrela}\\
		&\nabla_{\mu}\left\{\mathcal{I}\left[k^{\mu}-2\epsilon |z_1(0)|^2 \partial^{\mu}\gamma-\epsilon (\ell+\sigma) B^{\mu}   \right]\right\} =0,\label{eq-rtran}
\end{align}
where $B_{\mu}=i \bar{m}^{\alpha} \stackrel{h}{\nabla}_{\mu} m_{\alpha}$ represents the Berry connection projected onto the null tetrad $\{k_{\alpha}, n_{\alpha}, m_{\alpha}, \bar{m}_{\alpha}\}$, and $\gamma(\tau) = \int_{\tau_0}^{\tau} \mathrm{d} \tau k^\mu B_\mu$ denotes the Berry phase. 

Consequently, we derive an effective Hamilton-Jacobi equation from Eqs. \eqref{eq-rrela} and \eqref{eq-rtran}, as follows:
\begin{align}\label{eq:eff_HJ}
H(x, \nabla \tilde{S}) & =\frac{1}{2} g^{\mu \nu} k_\mu k_\nu +2\epsilon|z_1(0)|^2 k^\mu \partial_{\mu}\gamma -\epsilon (\sigma+\ell) k^\mu B_\mu \nonumber\\
		& =\frac{1}{2} g^{\mu \nu} \nabla_{\mu} \tilde{S} \nabla_{\nu} \tilde{S}-\epsilon (\sigma+\ell) g^{\mu \nu} B_{\mu} \nabla_{\nu} \tilde{S},
\end{align}
where $\Tilde{S}(x) = S(x) + 2 \epsilon |z_1(0)|^2 \gamma$ signifies the total phase, representing the phase factor for the WKB solution, $\mathcal{A}_\alpha = \mathrm{Re} (\sqrt{\mathcal{I}}m_a e^{i\ell \phi} e^{2i\gamma}e^{iS/\epsilon})$. 

\subsection{Effective ray equations for vortex light}

To derive the effective ray equations, it is convenient to express the Hamiltonian $H(x, \nabla \tilde{S})$ on the cotangent  bundle $T^*M$ as follows:
\begin{equation}\label{Hamiltonian}
	H(x, p)=\frac{1}{2} g^{\mu \nu} p_{\mu} p_{\nu}-\epsilon (\sigma+\ell) g^{\mu \nu} p_{\mu} B_{\nu}(x, p),
\end{equation}
where $p_\mu$ denotes the momentum of light. Following this, the effective ray equations for vortex light can be derived from the corresponding Hamilton's equations:
\begin{align}
	\dot{x}^\mu &= g^{\mu \nu} p_\nu - \epsilon (\sigma+\ell) \left( B^\mu + p^\alpha \vnabla{}^\mu B_\alpha \right), \label{eq:EOM_1_x}\\
	\dot{p}_\mu &=  -\frac{1}{2} \partial_\mu g^{\alpha \beta} p_\alpha p_\beta + \epsilon (\sigma+\ell) p_\alpha \left( \partial_\mu g^{\alpha \beta} B_\beta + g^{\alpha \beta} \partial_\mu B_\beta \right). \label{eq:EOM_1_p}
\end{align}
Here, the notation $\dot{x}^\mu$ and $\dot{p}_\mu$ represents differentiation with respect to $d/d\tau$, which denotes the world line parameter for vortex light.

Notably, the Berry connection $B_\mu$ is gauge-dependent. This implies that both the Hamiltonian $H(x,\nabla \widetilde{S})$ and its associated Hamilton's equations \eqref{eq:EOM_1_x} and \eqref{eq:EOM_1_p} are influenced by the choice of null tetrad $m_\alpha$ and $\bar{m}_\alpha$. To address this issue and derive a set of gauge-independent effective ray equations, we introduce a new set of noncanonical coordinates, defined as:
\begin{align} \label{eq:coord}
	X^\mu &= x^\mu + i \epsilon (\sigma+\ell) \bar{m}^{\alpha} \vnabla{}^\mu {m}_\alpha, \\
	P_\mu &= p_\mu - i \epsilon (\sigma+\ell) \bar{m}^{\alpha} \nabla_\mu {m}_\alpha. \label{eq:coord1}
\end{align}
Within this new coordinate system $(X,P)$, we derive a modified Hamiltonian from (\ref{Hamiltonian}):
\begin{equation}\label{eq:coordinvham}
	H'(X, P) = \frac{1}{2}g^{\mu \nu}(X) P_\mu P_\nu.
\end{equation}

Following detailed calculations (refer to Appendix \ref{appendix3}), the effective ray equations in these new coordinates are:
\begin{align}
	\dot{X}^\mu \, &= \, P^\mu + \epsilon (\sigma+\ell) P^\nu \left( F_{p x} \right)\indices{_\nu^\mu}+ \epsilon (\sigma+\ell) \Gamma^\alpha_{\beta \nu} P_\alpha P^\beta \left( F_{p p}\right)^{\nu \mu} ,\label{eq:Xdot}  \\
   \dot{P}_\mu \, &= \, \Gamma^\alpha_{\beta \mu} P_\alpha P^\beta - \epsilon (\sigma+\ell) P^\nu \left( F_{x x} \right)_{\nu \mu}- \epsilon (\sigma+\ell) \Gamma^\alpha_{\beta \nu} P_\alpha P^\beta \left( F_{x p} \right)\indices{^\nu_\mu} , \label{eq:Pdot}
\end{align}
where $\left( F_{p x} \right)\indices{_\nu^\mu}$, $\left( F_{p p}\right)^{\nu \mu}$, $\left( F_{x x} \right)_{\nu \mu}$ and $\left( F_{x p} \right)\indices{^\nu_\mu}$ represent the Berry curvature terms. In these equations, $\sigma+\ell$ denotes the total angular momentum of vortex light along its propagation direction. As $\sigma+\ell=0$, these equations reduce to the standard null geodesic equations in general relativity.

Moreover, the terms proportional to $\sigma+\ell$ highlight that the propagation of vortex light is influenced by both its spin and IOAM, essentially by its total angular momentum. Specifically, for $\ell=0$ and $\sigma \neq 0$, indicating spin-polarized light without exhibiting vortex states, these equations align with those provided by Oancea et al. \cite{oancea2020gravitational}, demonstrating the SHE in curved spacetime. Conversely, for $\sigma=0$ and $\ell \neq 0$, showing vortex states without spin polarization, the equations highlight the influence of IOAM, known as the OHE. Additionally, up to the first order in $\epsilon$, only the terms proportional to the total angular momentum $\sigma+\ell$ are present. This suggests that under the WKB approximation, the effect of IOAM on light propagation in curved spacetime might be equivalent to that of spin.

\section{Orbital Hall effect in optical waveguides}\label{material candidates}

As introduced in Section \ref{introduction}, we can effectively model the behavior of light in inhomogeneous media by treating the media as curved spacetimes. This modeling suggests that Equations \eqref{eq:Xdot} and \eqref{eq:Pdot} are effective for describing the propagation of vortex light within optical media. Building on this foundation, this section explores the OHE of vortex light as it propagates through an optical waveguide, serving as an example to investigate the effects of IOAM.

A perfect impedance-matched optical medium can be effectively modeled by a curved spacetime, where the metric $g_{\mu\nu}$ is described as follows \cite{PhysRev.118.1396,carini1992phase,wu2022testing,chanda2019jacobi,gibbons2019gravitational,fernandez2016anisotropic}:
\begin{align}\label{tensorial permittivity,tensorial permeability}
	\sqrt{-g} \frac{g^{i j}}{g_{00}}=-\epsilon^{i j}=-\mu^{i j}, \quad \frac{g_{0 i}}{g_{00}}=-\alpha_{i},
\end{align}
where $\epsilon_{ij}$ and $\mu_{ij}$ represent the tensorial permittivity and permeability, respectively, and $\alpha_i$ denotes the magnetoelectric coupling vector. Specifically, we define the medium's refractive index as $n=n(t,x,y,z)$ and its four-velocity by $u_\alpha$. In this case, the metric for the effective curved spacetime simplifies to \cite{PhysRevD.97.065001,gordon1923lichtfortpflanzung,PhysRevD.105.104061}:
\begin{equation}
	g_{\alpha \beta}=\eta_{\alpha \beta}+(1-n^{-2}) u_{\alpha} u_{\beta},
\end{equation} 
where $\eta_{\mu \nu} = \text{diag}\{-1,1,1,1\}$ is the Minkowski metric. Assuming a stationary medium with $\partial_t n = 0$ and setting $u^\alpha = (1,0,0,0)$, this above metric can be simlified to
\begin{equation}\label{metric1}
	g_{\alpha \beta}=\eta_{\alpha \beta}+(1-n^{-2}) \delta_{0\alpha}\delta_{0\beta}.
\end{equation}

\subsection{Effective ray equations for vortex light in the stationary medium}

To calculate the Berry curvatures in the effective ray equations, we can select the tetrad $\{t^\alpha,p^\alpha,v^\alpha,w^\alpha\}$, which satisfies the conditions:
\begin{equation}\label{eq-4tetrad}
	\begin{array}{c}
		t_{\alpha} t^{\alpha}=-1, \quad p_{\alpha} p^{\alpha}=0, \quad t_{\alpha} p^{\alpha}=-\epsilon \omega, \quad v_{\alpha} v^{\alpha}=w_{\alpha} w^{\alpha}=1, \\
		t_{\alpha} v^{\alpha}=t_{\alpha} w_{\alpha}=p_{\alpha} v^{\alpha}=p_{\alpha} w^{\alpha}=v_{\alpha} w^{\alpha}=0,
	\end{array}
\end{equation}
where, $t^\alpha$ represents the observer's velocity vector field. The vectors $v^\alpha$ and $w^\alpha$ are real spacelike and satisfy:
\begin{equation}
	m^{\alpha}=\frac{1}{\sqrt{2}}\left(v^{\alpha}+i w^{\alpha}\right), \quad \bar{m}^{\alpha}=\frac{1}{\sqrt{2}}\left(v^{\alpha}-i w^{\alpha}\right).
\end{equation}
The Berry curvatures are thus expressed as:
\begin{align}
		&\left({F_{p p}}\right)^{\nu \mu} =   \vnabla{}^\nu v^\alpha  \vnabla{}^\mu w_\alpha - \vnabla{}^\mu v^\alpha \vnabla{}^\nu w_\alpha ,\\
		&\left({F_{xx}}\right)_{\nu \mu}=\nabla_{\nu} v^{\alpha} \nabla_{\mu} w_{\alpha}-\nabla_{\mu} v^{\alpha} \nabla_{\nu} w_{\alpha},\\
		&\left({F_{p x}}\right)\indices{_\nu^\mu}=\nabla_{\nu} v^{\alpha} \vnabla{}^\mu w_{\alpha}-\vnabla{}^{\mu} v^{\alpha} \nabla_{\nu} w_{\alpha}.
\end{align}

As outlined in \cite{PhysRevA.92.043805}, a specific set of polarization vectors can be conveniently constructed by introducing an orthonormal tetrad $(e_a)^\alpha$, with $(e_0)^\alpha=t^\alpha$. Consequently, the vectors $p^\alpha$, $v^\alpha$ and $w^\alpha$ can be given by $p^\alpha=P^a(e_a)^\alpha$, $v^\alpha=V^a(e_a)^\alpha$ and $w^\alpha=W^a(e_a)^\alpha$, respectively. The components of these vectors are specified as follows:
\begin{align}
	P^{a}=\left(\begin{array}{l}
		P^{0} \\
		P^{1} \\
		P^{2} \\
		P^{3}
	\end{array}\right), \quad V^{a}=\frac{1}{P_{p}}\left(\begin{array}{c}
		0 \\
		-P^{2} \\
		P^{1} \\
		0
	\end{array}\right), \quad
	W^{a}=\frac{1}{P_{p} P_s}\left(\begin{array}{c}
		0 \\
		P^{1} P^{3} \\
		P^{2} P^{3} \\
		-\left(P_{p}\right)^{2}
	\end{array}\right),
\end{align}
where
\begin{align}
	P_{p} & =\sqrt{\left(P^{1}\right)^{2}+\left(P^{2}\right)^{2}}, \\
	P_s& =\sqrt{\left(P^{1}\right)^{2}+\left(P^{2}\right)^{2}+\left(P^{3}\right)^{2}}.
\end{align}

In this stationary medium as described by the metric \eqref{metric1}, we adopt an orthonormal tetrad as follows:
\begin{equation}\label{tetrad}
	e_{0}=n \partial_{t}, \quad e_{1}=\partial_{x}, \quad e_{2}=\partial_{y}, \quad e_{3}=\partial_{z}.
\end{equation}
The tetrad components $(e_a)^\alpha$ are determined by $e_a=(e_a)^\alpha\partial_\alpha$. In this case, the velocity vector field of the observer can be set as $t^\alpha = (e_0)^\alpha=(n,0,0,0)$. Outside the optical medium where $n = 1$, this timelike observer reduces to the Minkowski observer $\partial_t$, which is generally assumed in optics when studying the OHE of light. 

Within the framework of the tetrad defined by Eq. \eqref{tetrad}, the terms in the ray equations \eqref{eq:Xdot} and \eqref{eq:Pdot} are specified as follows:
\begin{align}
	\Gamma_{\beta \nu}^{\alpha} P_{\alpha} P^{\beta}\left(F_{pp}\right)^{\nu \mu} = \begin{pmatrix}0\\ \frac{\mathbf{\dot{P}} \times \mathbf{P}}{P^3}\end{pmatrix},
	\quad \Gamma_{\beta \mu}^{\alpha} P_{\alpha} P^{\beta} = nP_{0}^{2} \partial_{\mu} n
\end{align}
where $P=\sqrt{P_1^2+P_2^2+P_3^2}$, and all other terms involving Berry curvatures are zero in this case. Consequently, the ray equations for vortex light simplify to:
\begin{align}\label{eq-ray}
	\frac{dX^0}{d\tau}=P^0,\quad \frac{d\mathbf{X}}{d\tau} =\mathbf{P}+\epsilon (\sigma+\ell)\frac{\mathbf{\dot{P}} \times \mathbf{P}}{P^3},\quad \frac{d P_i}{d\tau} = nP_0^2\partial_i n.
\end{align}
Here, $\mathbf{X}=\{X^i\}$ and $\mathbf{P}=\{P_i\}$ represent the spatial components of the position $X^\mu$ and momentum $P_\mu$, respectively. Given the stationary medium condition $\partial_t n=0$, it follows that $d P_0/d\tau=0$, a detail not explicitly presented in the preceding equations.

The Hamiltonian for vortex light, given by $H=\frac{1}{2}g^{\alpha\beta}P_\alpha P_\beta=0$, together with the metric $g_{\alpha\beta}=\eta_{\alpha\beta}+(1-n^{-2})\delta_{0\alpha}\delta_{0\beta}$, yields the relation:
\begin{equation}
n^{2} P_{0}^{2} =P_{1}^{2}+P_{2}^{2}+P_{3}^{2}=P^{2}.
\end{equation}
Consequently, the ray equations in Eq. \eqref{eq-ray} can be rewritten as:
\begin{align}
	\frac{dX^0}{d\tau}=nP,\quad \frac{d\mathbf{X}}{d\tau} =\mathbf{P}+\epsilon (\sigma+\ell)\frac{\mathbf{\dot{P}} \times \mathbf{P}}{P^3},\quad \frac{d P_i}{d\tau} = \frac{P^{2}}{n} \partial_{i} n,	
\end{align}
where we have used the relation $P_0=-P/n$. Given $dX^0/d\tau=nP$, the ray equations simplify further to: 
\begin{align}\label{eq-rayt}
	\frac{d \mathbf{X}}{d T}=\frac{\mathbf{P}}{n P}+\frac{\epsilon (\sigma+\ell)}{P^3} \frac{d \mathbf{P}}{d T} \times \mathbf{P},\quad \frac{d P_i}{d T}=\frac{P}{n^{2}} \partial_i n.
\end{align}

This above equation is consistent with the effective ray equations for spin-polarized light in Ref. \cite{PhysRevD.105.104061}. In several studies \cite{ling2017recent,bliokh2009geometrodynamics,Bliokh_2006,PhysRevA.92.043805}, the effective ray equations for spin-polarized light are also presented as:
\begin{align}
\frac{d\mathbf{X}}{dT}=\frac{\boldsymbol{\rho}}{\rho} + \frac{\epsilon \sigma}{\rho^3} \frac{d\boldsymbol{\rho}}{dT}  \times \boldsymbol{\rho},\quad \frac{d\boldsymbol{\rho}}{dT}=\nabla n,
\end{align}
where $\boldsymbol{\rho}\equiv n \mathbf{P} /P$ and $\rho=|\boldsymbol{\rho}|$. Substituting $\mathbf{P}$ with $\boldsymbol{\rho}$ in Eq. \eqref{eq-rayt} aligns the effective ray equations for vortex light with those for spin-polarized light. This alignment suggests that the intrinsic orbital angular momentum of vortex light, when interacting with a medium, can produce effects analogous to those associated with spin.

As outlined in Eq. \eqref{eq-4tetrad}, the WKB parameter $\epsilon$ is connected to the observed frequency of light, $\omega$, by the relation $\epsilon = -t^\alpha p_\alpha / \omega$. By considering the relationships $p^\alpha = P^a (e_a)^\alpha$ and $t^\alpha = (n,0,0,0)$, we can express $\epsilon$ in terms of $\omega$ as $\epsilon = -(e_0)^\alpha P^a (e_a)^\beta g_{\alpha\beta} / \omega$. From Eqs. \eqref{metric1} and \eqref{tetrad}, this relationship simplifies to $\epsilon = -P_0 / \omega$. It is noteworthy that $P_0 = -P/n$ remains constant, as indicated by $dP_0 / d\tau = 0$. Consequently, $\epsilon = P/n\omega$ can be interpreted as the wavelength of vortex light, adopting the natural units where $c = \hbar = G = 1$.

\subsection{Ray trajectories for vortex light in a specific optical waveguide}

As an example, we consider a waveguide with its refractive index given by:
\begin{equation}\label{waveguide with a thick adiabatic gradient}
	n(X,Y)=n_{0}+a /\left(1+\left(\sqrt{X^{2}+Y^{2}} / r_{c}\right)^{8}\right),
\end{equation}
where $n_0=1.37$, and the parameters $a$ and $r_c$ depend on the waveguide's characteristics. This design enables the simulation of gravitational lensing, including the formation of Einstein rings, following the approach in \cite{sheng2016wavefront}. Using this index profile, we numerically solve the effective ray equations (\ref{eq-rayt}) using Mathematica, setting $r_c=10$ and $a=1$. For these simulations, the vortex light's wavelength, defined as $\lambda=1/\omega$, is fixed at  $\lambda=0.04$, with its spin parameter set to $\sigma=0$.

\begin{figure}
	\includegraphics[width=0.7\textwidth]{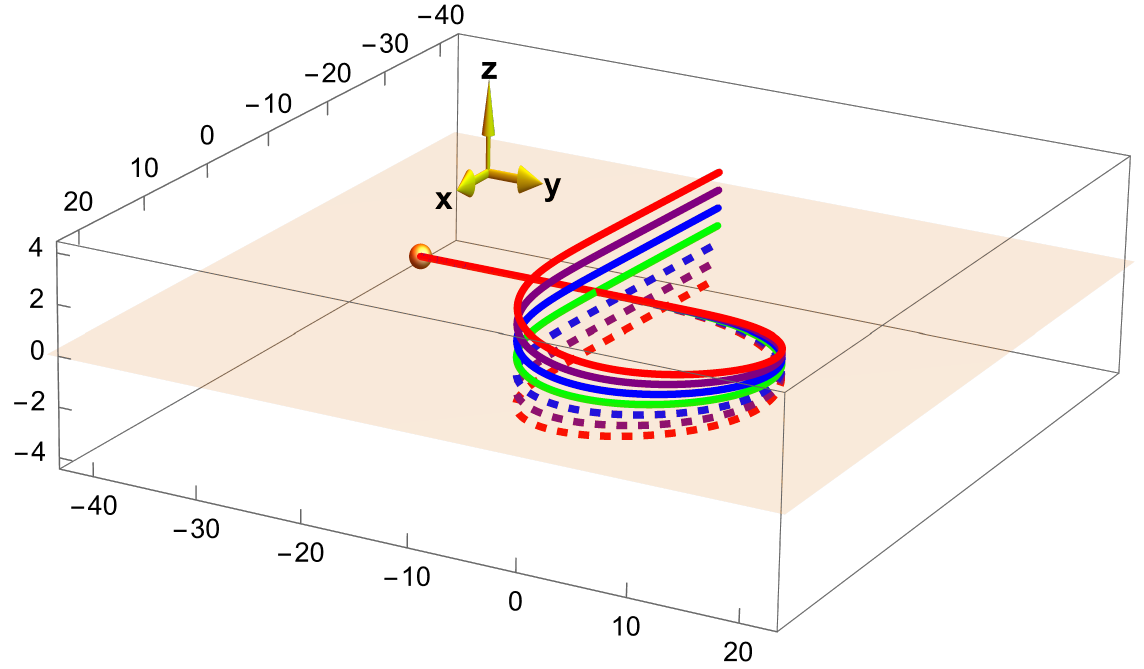}\\
	\includegraphics[width=0.6\textwidth]{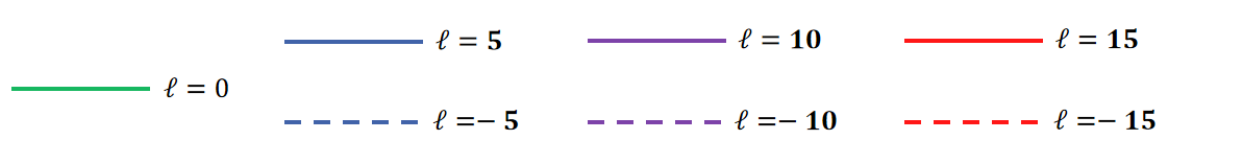}
	\caption{Trajectories of vortex light beams within the optical waveguide. The small orange ball represents the initial position of the vortex light, $\mathbf{X}(0)= (-13.4,-30,0)$. The initial momentum is $\mathbf{P}(0)=(0,1,0)$.}\label{fig:diagram}
\end{figure}

Fig. \ref{fig:diagram} shows the trajectories of vortex light with varying orbital angular momentum $\ell$ in three-dimensional space. The initial position and momentum of the vortex light are set to $\mathbf{X}(0)= (-13.4,-30,0)$ and $\mathbf{P}(0)=(0,1,0)$, respectively. A key observation is that the intrinsic orbital angular momentum causes the vortex light to separate along the $Z$-axis, which is perpendicular to the propagation plane of light without angular momentum. This separation, termed the OHE, closely resembles the SHE in behavior.

The refractive index $n(X,Y)$ of the waveguide is independent of the $Z$-axis, leading to $dP_Z/dT=0$ as described by Eq. \eqref{eq-rayt}. With the initial momentum $\mathbf{P}(0)=(0,1,0)$, it follows that $P_Z(T)$ remains zero throughout the propagation of vortex light in the waveguide. Consequently, the ray trajectory of the vortex light along the $Z$-axis is governed by
\begin{equation}\label{eq-dzt}
	\frac{dZ}{dT} = \frac{\lambda\ell (P_Y\partial_X n-P_X\partial_Y n)}{n^3 P},
\end{equation}
where $\epsilon=P/n\omega$ and $\lambda=1/\omega$ are taken into account. This equation implies that the ray trajectory along the  $Z$-axis is directly proportional to the intrinsic orbital angular momentum $\ell$ of the vortex light. This proportionality is consistent with the results shown in Fig. \ref{fig:diagram}.

According to Eq. \eqref{waveguide with a thick adiabatic gradient}, the refractive index gradient, $\nabla n$, varies with the position's distance from the origin. Specifically, for vortex light located at $Y$-axis, $\partial_X n$ becomes zero due to $X=0$. Given the initial momentum $\mathbf{P}(0)=(0,1,0)$, vortex light possessing varying IOAM does not separate along the $Z$-axis, as $P_X(T)=0$ persists during propagation in the waveguide, as shown by Eq. \eqref{eq-rayt}. Therefore, the ray trajectory of vortex light along the $Z$-axis depends on the initial position's distance from the $Y$-axis.

\begin{figure}
	\subfigure[$X_0=-13.6$]{\includegraphics[width=0.3\textwidth]{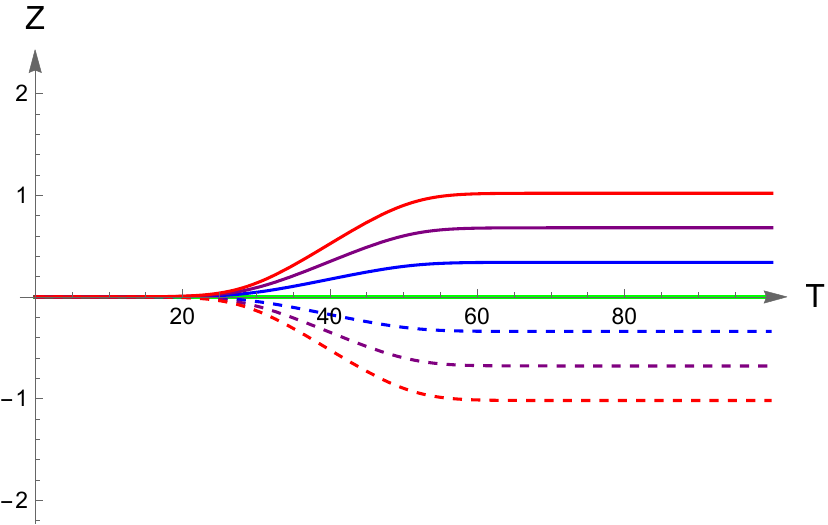}}
	\subfigure[$X_0=-13.4$]{\includegraphics[width=0.3\textwidth]{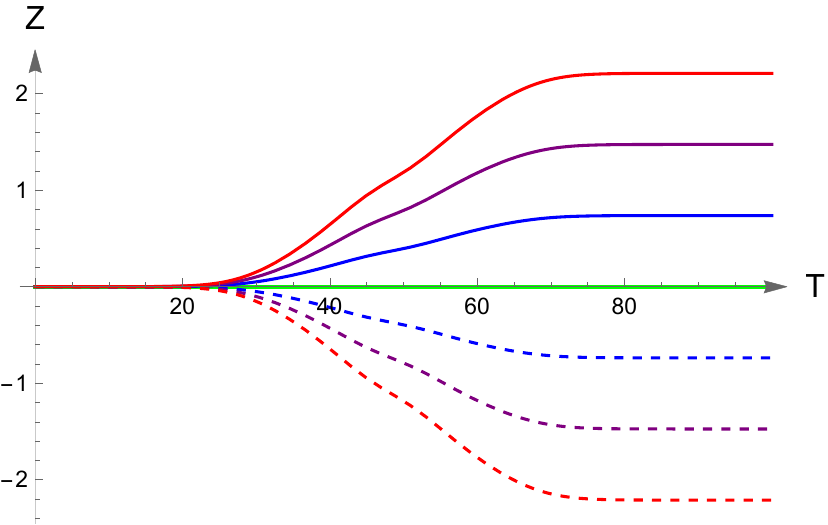}}
	\subfigure[$X_0=-11.0$]{\includegraphics[width=0.3\textwidth]{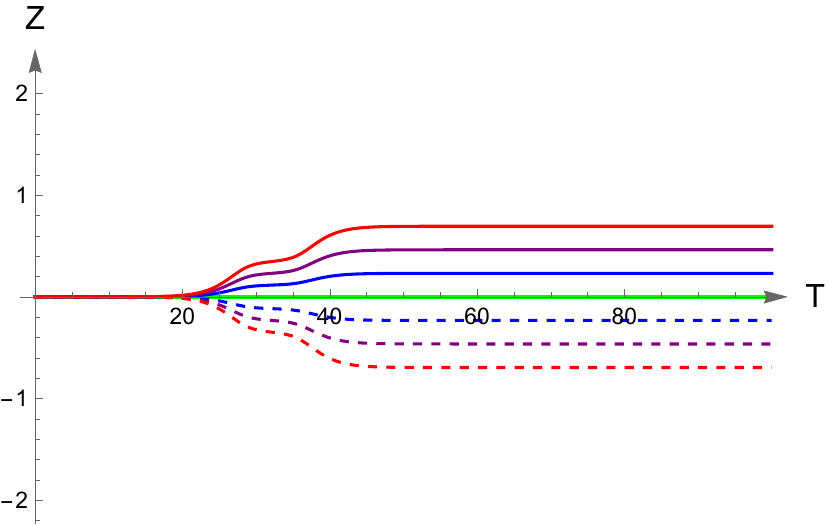}}\\
	\subfigure[$X_0=11.0$]{\includegraphics[width=0.3\textwidth]{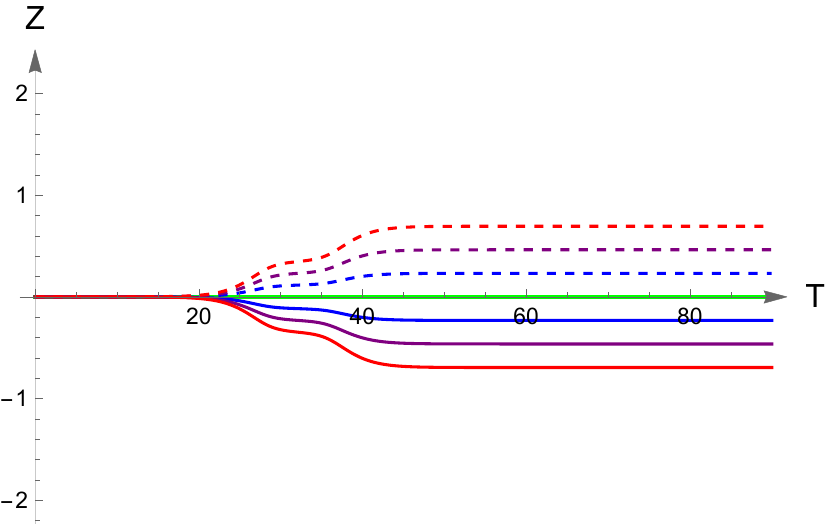}}
	\subfigure[$X_0=13.4$]{\includegraphics[width=0.3\textwidth]{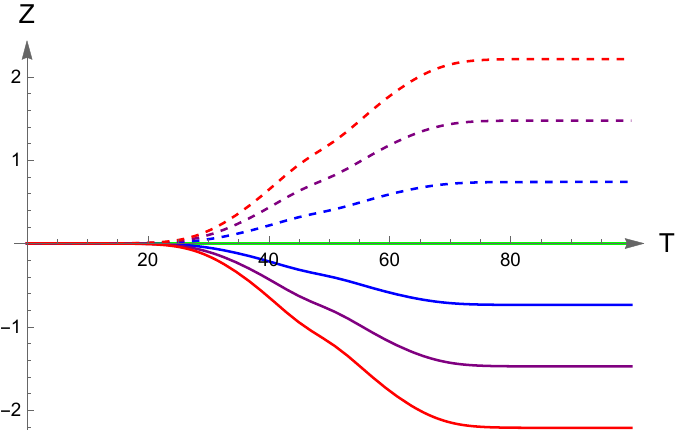}}
	\subfigure[$X_0=13.6$]{\includegraphics[width=0.3\textwidth]{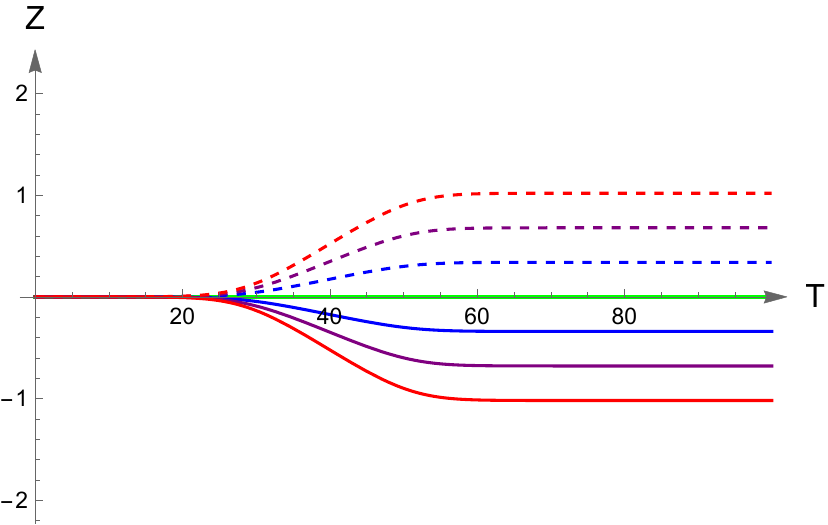}}
	\caption{Trajectories of vortex light beams along the $Z$-axis with different initial positions $\mathbf{X}(0)=(X_0,-30,0)$.}\label{fig.res2}
\end{figure}

Fig. \ref{fig.res2} shows the trajectories of vortex light beams along the $Z$-axis for different initial positions $\mathbf{X}(0)=(X_0,-30,0)$. As the initial position moves from infinity towards the $Y$-axis, the separation between vortex light beams with opposite orbital angular momentum initially increases, subsequently decreases, and ultimately vanishes when the beams are directly on the $Y$-axis. This observation is consistent with the preceding discussion. Furthermore, as $X_0<0$, vortex light beams with positive intrinsic orbital angular momentum ($\ell>0$) diverge along the $Z$-axis in the positive direction. Conversely, for $X_0>0$, the divergence occurs in the negative direction.

\begin{figure}
\subfigure[$X_0=-13.6$]{\includegraphics[width=0.3\textwidth]{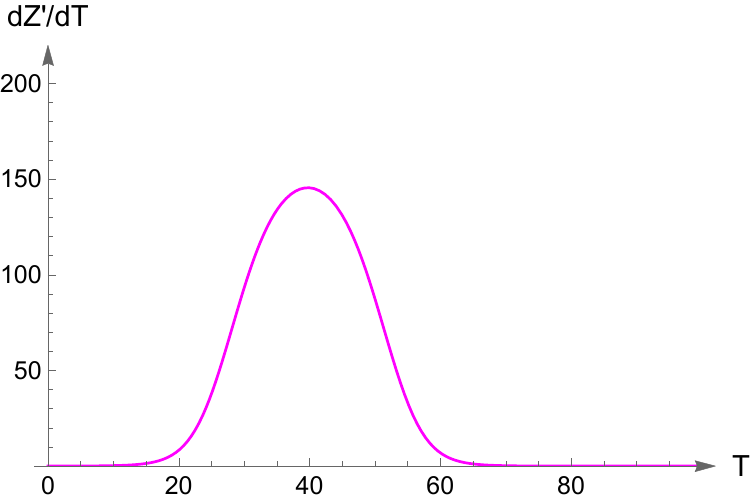}}
\subfigure[$X_0=-13.4$]{\includegraphics[width=0.3\textwidth]{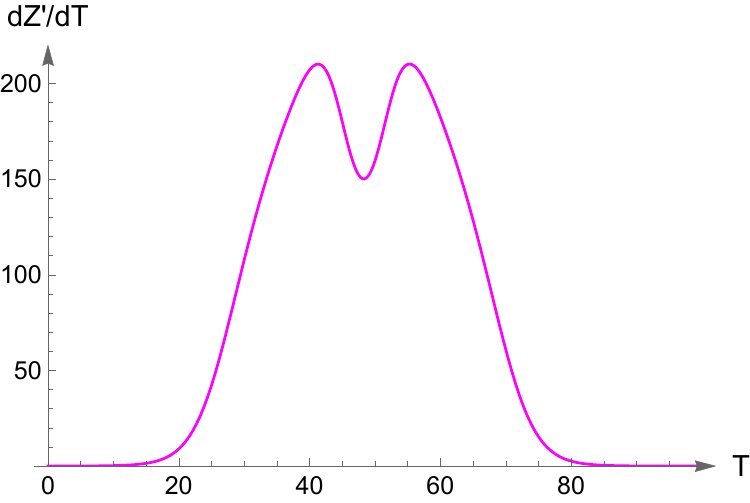}}
\subfigure[$X_0=-11.0$]{\includegraphics[width=0.3\textwidth]{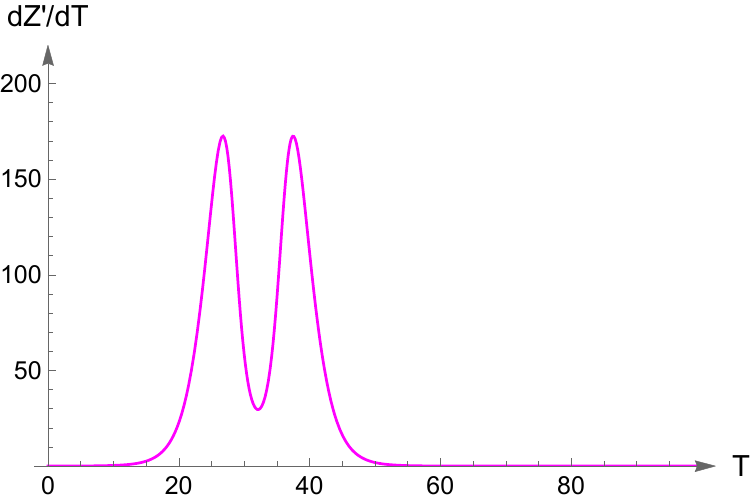}}\\
\subfigure[$X_0=11.0$]{\includegraphics[width=0.3\textwidth]{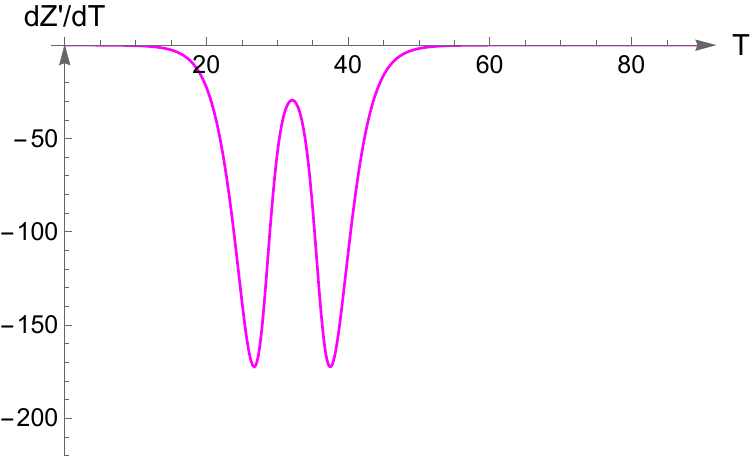}}
\subfigure[$X_0=13.4$]{\includegraphics[width=0.3\textwidth]{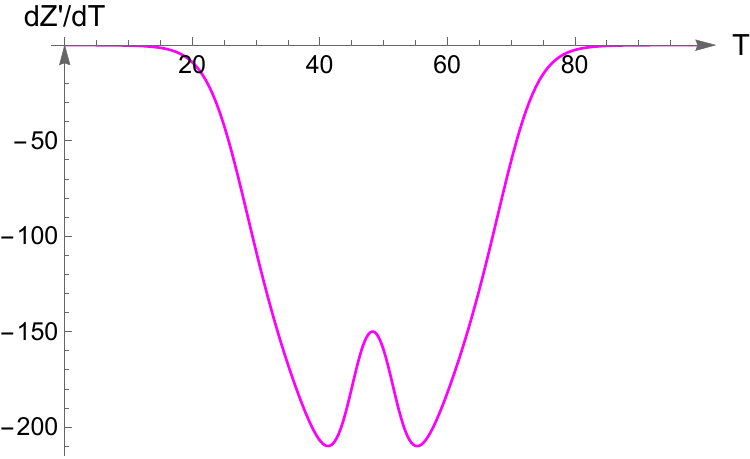}}
\subfigure[$X_0=13.6$]{\includegraphics[width=0.3\textwidth]{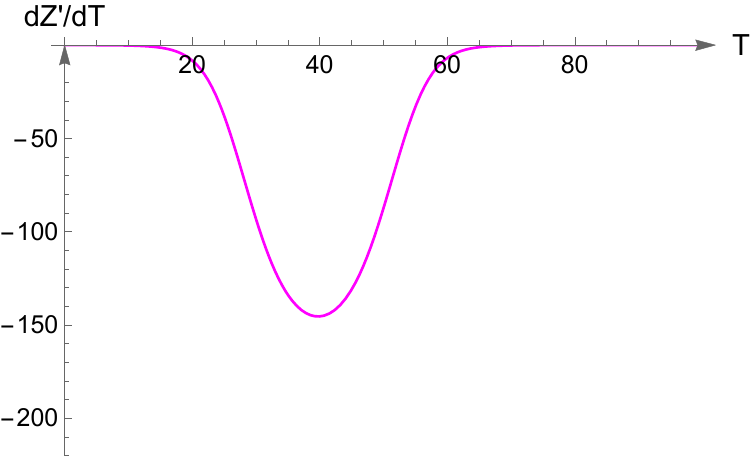}}
	\caption{Rate of change $dZ/dT$  versus initial position $\mathbf{X}(0)=(X_0,-30,0)$ for vortex light with intrinsic orbital angular momentum $\ell=5$.}\label{fig.res3}
\end{figure}

Additionally, when $X_0=\pm 11.0$, ray trajectories along the $Z$-axis exhibit two distinct inflection points, indicating a near-zero rate of change within this range. Fig. \ref{fig.res3} shows the variation of $dZ/dT$ as the ray's initial position shifts from infinity towards the $Y$-axis, revealing two peaks in $dZ/dT$ for $|X_0|<=13.4$. This pattern aligns with the aforementioned behavior, suggesting that the trajectory of vortex light in optical waveguides is influenced by both its intrinsic orbital angular momentum and initial position. Consequently, the intrinsic orbital angular momentum of vortex light could result in novel and complex behaviors during its propagation in optical waveguides.

\section{conclusion and discussion}\label{conclusions}

In this study, we derive effective ray equations for vortex light propagating in curved spacetime and optical media by applying the WKB approximation to the covariant Maxwell equations. The dynamics of vortex light are captured by the effective Hamiltonian:
\begin{equation}
	H(x, p)=\frac{1}{2} g^{\mu \nu} p_{\mu} p_{\nu}-\epsilon (\sigma+\ell) g^{\mu \nu} p_{\mu} B_{\nu}(x, p).
\end{equation}
The corresponding effective ray equations are:
\begin{align}
	\begin{split}\dot{X}^\mu \, &= \, P^\mu + \epsilon (\sigma+\ell) P^\nu \left( F_{p x} \right)\indices{_\nu^\mu}+ \epsilon (\sigma+\ell) \Gamma^\alpha_{\beta \nu} P_\alpha P^\beta \left( F_{p p}\right)^{\nu \mu} ,\end{split} \label{eq:Xdot2} \\
	\begin{split}\dot{P}_\mu \, &= \, \Gamma^\alpha_{\beta \mu} P_\alpha P^\beta - \epsilon (\sigma+\ell) P^\nu \left( F_{x x} \right)_{\nu \mu}- \epsilon (\sigma+\ell) \Gamma^\alpha_{\beta \nu} P_\alpha P^\beta \left( F_{x p} \right)\indices{^\nu_\mu} .\end{split} \label{eq:Pdot2}
\end{align}

According to these effective ray equations, the IOAM of vortex light influences its propagation through optical media and curved spacetimes similarly to spin. This suggests that light propagation is determined by its total angular momentum, regardless of whether it stems from spin or orbital contributions. Thus, IOAM could induce phenomena akin to those associated with spin, highlighting the unified role of angular momentum in light propagation.

To investigate the effects of IOAM on light propagation, we analyze the trajectory of vortex light in an optical waveguide characterized by the refractive index $n(x, y)=n_{0}+a /\left(1+(x^{2}+y^{2})^4 / r_{c}^{8}\right)$. By solving the aforementioned effective ray equations, we find that vortex beams possessing $\ell$ diverge perpendicularly to the propagation plane, denoted as the $Z$-axis. This divergence, known as the OHE, is directly proportional to the IOAM. Our findings further reveal that the initial position of the light beams significantly influences this divergence. As vortex light travels within the waveguide, its divergence along the $Z$-axis shows two distinct inflection points, indicating a near-zero rate of change in this region. These results suggest that the IOAM of vortex light might induce novel phenomena in optical media and curved spacetimes.

Additionally, in deriving the effective ray equations, we adopted a preliminary finding: the combined effects of spin and IOAM's Berry connections, for vortex light propagating along a curved ray, can be represented as a unified Berry connection. This finding has been supported and validated by several studies \cite{Bliokh_2006,PhysRevA.97.033843,PhysRevA.92.043805,DASGUPTA200691,PhysRevA.46.5199,PhysRevE.70.026605}. However, recent research investigating the OHE of vortex particles by directly solving their equations of motion \cite{lian2024motion,qiu2024orbital} indicates that IOAM's effects might slightly diverge from those of spin. The difference between these two approaches highlights the need for more in-depth research into the roles of spin and IOAM in light propagation. Observing the OHE for vortex light propagating within optical waveguides could offer insights to bridge these methodological differences, deepening our comprehension of how spin and IOAM influence light propagation.

\appendix

\section{The effective Hamilton-Jacobi system for vortex light}\label{appendix2}

The second term in Eqs. \eqref{seq-rela} and \eqref{seq-trans}, given as $ia_{0}^{* \alpha} \nabla_{\mu} a_{0 \alpha}$, is commonly interpreted as the Berry connection associated with the spin of light. Similarly, the third term in these equations, $i e^{-i \ell \phi} \nabla_{\mu}e^{i \ell \phi}$, shares the spine form as the second term and is typically identified as the Berry connection corresponding to the IOAM for vortex light. Research indicates that for vortex light propagating along a curved ray, the combined effect of these two Berry connections can be represented as a unified Berry connection \cite{Bliokh_2006,PhysRevA.97.033843,PhysRevA.92.043805,DASGUPTA200691,PhysRevA.46.5199,PhysRevE.70.026605}:
\begin{equation}
	\mathcal{B_{\mu}}^{\sigma \sigma}+\mathcal{B_{\mu}}^{\ell \ell}=(\sigma+\ell) \mathcal{B}'_\mu,
\end{equation}
where
\begin{equation}\label{Berry connection1}
	\mathcal{B_{\mu}}^{\sigma \sigma}=i {a_0}^{*\alpha} \nabla_\mu {a_0}_\alpha, \quad \mathcal{B_{\mu}}^{\ell \ell}=i e^{-i \ell \phi} \nabla_{\mu}e^{i \ell \phi}.
\end{equation}

As described by Eqs. \eqref{seq-rela} and \eqref{seq-trans}, the effective Hamilton-Jacobi equations for vortex light are:
\begin{align}
	&\frac{1}{2} g^{\mu \nu} k_{\mu} k_{\nu} -i\epsilon k^{\mu} a'^{*\alpha}\nabla_\mu a'_{\alpha} = 0,\label{a-rela}\\
	&\nabla_{\mu}\left\{\mathcal{I}(k^{\mu}-i\epsilon g^{\mu \nu}a'^{*\alpha}\nabla_\mu a'_{\alpha})\right\} =0,\label{a-trans}
\end{align}
where $a'_\alpha=a_{0\alpha}e^{i\ell\phi}$. According to these equations, it is significantly that the propagation of vortex light is affected by its spin and IOAM. To detail these effects, we need to introduce a null tetrad $\left\{k_{\alpha}, n_{\alpha}, m_{\alpha}, \bar{m}_{\alpha}\right\}$, satisfying:
\begin{equation}\label{eq:NPtetrad}
	\begin{split}
		m_\alpha \bar{m}^\alpha = 1, \qquad k_\alpha n^\alpha = -1, \\
		k_\alpha k^\alpha = n_\alpha n^\alpha = m_\alpha m^\alpha = \bar{m}_\alpha \bar{m}^\alpha &= 0, \\
		k_\alpha m^\alpha = k_\alpha \bar{m}^\alpha = n_\alpha m^\alpha = n_\alpha \bar{m}^\alpha &= 0.
	\end{split}
\end{equation}

The polarization vector $a'_\alpha$ satisfies the orthogonality condition $k^{\alpha} a'_{\alpha}=0$, due to $k^{\alpha} a_{0\alpha}=0$. As indicated by \cite{oancea2020gravitational}, $a'_\alpha$ can be represented as
\begin{equation} \label{eq:a0_basis}
	a'_\alpha(x, k) = z_1(x) m_\alpha(x, k) + z_2(x) \bar{m}_\alpha(x, k) + z_3(x) k_\alpha(x),
\end{equation}
where  $z_1$, $z_2$, and $z_3$ are complex scalar functions representing the projection values. The complex scalar functions $z_1$ and $z_2$ satisfy $z_1^* z_1 + z_2^* z_2=1$, considering $a'^{*\alpha}a'_\alpha=1$. In this case, we can use $z_1$ and $z_2$ to construct a two-dimensional complex vector $z$, defined as
\begin{equation}
	z = \begin{pmatrix} z_1 \\ z_2 \end{pmatrix}.
\end{equation}
It is significant that $z$ satisfies $z^\dag z=1$, with $z^\dag = (z_1^*\ z_2^*)$. 

Since ${a_0}_\alpha$ is a unit complex covector, the scalar functions $z_1$ and  $z_2$ are  constrained by
\begin{equation}\label{constrain}
	\begin{split}
		1 &= z_1^* z_1 + z_2^* z_2 = z^\dag z , \\
		\sigma &= z_1^* z_1 - z_2^* z_2 = z^\dagger \sigma_3 z,
	\end{split}
\end{equation}
where $\sigma=0,\pm1$ represents spins decided and $\sigma_3$ is the third Pauli matrix, 
\begin{equation}\label{Pauli}
	\sigma_3 = \begin{pmatrix} 1 & 0 \\ 0 & -1 \end{pmatrix}.
\end{equation}

Following that, Eq.~\eqref{eq:transp_a0_2} can be simlified to: 
\begin{equation} \label{knz1}
	\begin{split}
		k^\mu \nabla_{\mu}z = i k^\mu (i \bar{m}^{\alpha} \stackrel{h}{\nabla}_{\mu} m_{\alpha} \sigma_3+ i e^{-i \ell \phi} \nabla_{\mu}e^{i \ell \phi} \mathbb{I}) z,\\
	\end{split}
\end{equation}
where $\mathbb{I}$ is the $2\times2 $ identity matrix and the notation $\stackrel{h}{\nabla}_{\mu}$ acting on the vector is the horizontal derivative defined as $\stackrel{h}{\nabla}_{\mu} u_{\alpha}=\frac{\partial}{\partial x^{\mu}} u_{\alpha}-\Gamma_{\alpha \mu}^{\sigma} u_{\sigma}+\Gamma_{\mu \rho}^{\sigma} p_{\sigma} \frac{\partial}{\partial p_{\rho}} u_{\alpha}$, which is the real 1-form extending to general relativity \cite{oancea2020gravitational,bliokh2009spin,PhysRevA.92.043805}. We now introduce another definition of Berry connection $B_{\mu}=B_{\mu}(x,k)\equiv i \bar{m}^{\alpha} \stackrel{h}{\nabla}_{\mu} m_{\alpha}$ \cite{oancea2020gravitational}. It is expressed under the basis $m_\alpha, \bar{m}_\alpha$ and features the spine information as the gauge potential $\mathcal{A_{\mu}}$. We require that the relations between the physical quantities we describe remain the spine under the selection of different coordinate bases, so the proportional relation $B_{\mu} \sim \mathcal{A_{\mu}}$ is always true. Then with Eqs.~\eqref{eq:NPtetrad} $\sim$ \eqref{Pauli}, the effective Hamilton-Jacobi system can be simplified to
\begin{equation}\label{4}
	\begin{split}
		\frac{1}{2} g^{\mu \nu} k_{\mu} k_{\nu}-\frac{i \epsilon}{2} k^{\mu}\left(z^{\dagger} \partial_{\mu} z-\partial_{\mu} z^{\dagger} z\right)-\epsilon (\ell+\sigma) k^{\mu} B_{\mu}=\mathcal{O}\left(\epsilon^{2}\right),\\
		\nabla_{\mu}\left\{\mathcal{I}\left[k^{\mu}-\frac{i \epsilon}{2} \left(z^{\dagger} \partial^{\mu} z-\partial_{\mu} z^{\dagger} z\right)-\epsilon (\ell+\sigma) B^{\mu}\right]\right\} 
		=\mathcal{O}\left(\epsilon^{2}\right).
	\end{split}
\end{equation} 
and the Eq.~\eqref{knz1} can be rewritten as
\begin{equation} \label{knz2}
	\begin{split}
		k^\mu \nabla_{\mu}z =i k^\mu B_{\mu}(\sigma_3+\mathbb{I})z,
	\end{split}
\end{equation}

Furthermore, if we restrict z to an affinely parametrized null geodesic  $\tau \mapsto x^{\mu}(\tau)$ , with  $\dot{x}^{\mu}=k^{\mu}$ and integrate along the worldline, Eq.~\eqref{knz2} can be written as
\begin{equation} \label{eq:z_dot}
	\dot z = i k^\mu B_\mu (\sigma_3+\mathbb{I}) z,
\end{equation}
where $\dot z = \dot x^\mu \nabla_\mu z$. Integrating along the worldline, we obtain
\begin{equation} \label{eq:z_integration}
	z(\tau) = \begin{pmatrix} e^{i 2\gamma(\tau)} & 0 \\ 0 & 1 \end{pmatrix} z(0),
\end{equation}
where $\gamma$ represents the Berry phase \cite{bliokh2009spin,PhysRevA.92.043805}
\begin{equation}\label{Berryphase}
	\gamma(\tau_1) = \int_{\tau_0}^{\tau_1} \mathrm{d} \tau k^\mu B_\mu.
\end{equation}
The evolution of the first component of $z$ over time is decided by twice the Berry phase, while that of the second component remains unchanged as the initial state.
With Eqs.~\eqref{eq:z_integration} and \eqref{Berryphase}, Eq.~\eqref{4} can be simplified to:
\begin{equation}
	\begin{split}
		\frac{1}{2} g^{\mu \nu} k_\mu k_\nu +2\epsilon |z_1(0)|^2 k^\mu \partial_{\mu}\gamma -\epsilon (\sigma+\ell) k^\mu B_\mu=\mathcal{O}(\epsilon^2),\\
		\nabla_{\mu}\left\{\mathcal{I}\left[k^{\mu}-2\epsilon |z_1(0)|^2 \partial^{\mu}\gamma-\epsilon (\ell+\sigma)  B^{\mu}   \right]\right\} 
		=\mathcal{O}\left(\epsilon^{2}\right).
	\end{split}
\end{equation}
\section{Derivation of the effective Hamilton's equations in new coordinates}\label{appendix3}

Start from the Hamiltonian in noncanonical coordinates:
\begin{equation}
	H'(X, P) = \frac{1}{2}g^{\mu \nu}(X) P_\mu P_\nu.
\end{equation}
The Hamilton’s equations with new variables can be written in a matrix form as:
\begin{equation}
	\binom{\dot{X}^{\mu}}{\dot{P}_{\mu}}=T^{\prime}\binom{\frac{\partial H^{\prime}}{\partial X^{\nu}}}{\frac{\partial H^{\prime}}{\partial P_{\nu}}}.
\end{equation}
$T'$ is the Poisson tensor in the new variables. Following the results in \cite{marsden2013introduction}, we obtain
\begin{equation}
	T^{\prime}=\left(\begin{array}{cc}
		\epsilon (\sigma+\ell) \left(F_{p p}\right)^{\nu \mu} & \delta_{\nu}^{ \mu}+\epsilon (\sigma+\ell) \left(F_{x p}\right)_{\nu}{ }^{\mu} \\
		-\delta_{\mu}^{\nu}-\epsilon (\sigma+\ell)  \left(F_{x p}\right)^{\nu}{ }_{\mu} & -\epsilon (\sigma+\ell) \left(F_{x x}\right)_{\nu \mu}
	\end{array}\right),
\end{equation}
where the Berry curvature terms are
\begin{equation} \label{eq:Berry_curvature}
	\begin{split}
		\left({F_{p p}}\right)^{\nu \mu} &= i \Big( \vnabla{}^\mu \bar{m}^\alpha  \vnabla{}^\nu m_\alpha - \vnabla{}^\nu \bar{m}^\alpha \vnabla{}^\mu m_\alpha  + \bar{m}^\alpha \vnabla{}^{[\mu} \vnabla{}^{\nu]} m_\alpha - m_\alpha \vnabla{}^{[\mu} \vnabla{}^{\nu]} \bar{m}^\alpha \Big), \\
		\left({F_{xx}}\right)_{\nu \mu} &= i \Big( \nabla_\mu \bar{m}^\alpha \nabla_\nu m_\alpha - \nabla_\nu \bar{m}^\alpha \nabla_\mu m_\alpha + \bar{m}^\alpha \nabla_{[\mu} \nabla_{\nu]} m_\alpha - m_\alpha \nabla_{[\mu} \nabla_{\nu]} \bar{m}^\alpha \Big),  \\
		\left({F_{p x}}\right)\indices{_\nu^\mu} &= -\left({F_{x p}}\right)\indices{^\mu_\nu}
		= i \left( \vnabla{}^\mu \bar{m}^\alpha \nabla_\nu m_\alpha - \nabla_\nu \bar{m}^\alpha \vnabla{}^\mu m_\alpha \right).
	\end{split}
\end{equation}
The Poisson tensor in noncanonical coordinates $T'$ satisfies the Jacobi identity, since it is a covariant quantity obtained from the Poisson tensor $T$ in canonical coordinates
\begin{equation}
	T=\left(\begin{array}{cc}
		0 & \delta_{\nu}^{\mu} \\
		-\delta_{\mu}^{\nu} & 0
	\end{array}\right)
\end{equation}
through a change of variables on the cotangent bundle \cite{marsden2013introduction,oancea2020gravitational}.
Finally the effective ray equations can be obtained:
\begin{align}
	\begin{split}\dot{X}^\mu \, &= \, P^\mu + \epsilon (\sigma+\ell) P^\nu \left( F_{p x} \right)\indices{_\nu^\mu}+ \epsilon (\sigma+\ell) \Gamma^\alpha_{\beta \nu} P_\alpha P^\beta \left( F_{p p}\right)^{\nu \mu} ,\end{split} \\
	\begin{split}\dot{P}_\mu \, &= \, \Gamma^\alpha_{\beta \mu} P_\alpha P^\beta - \epsilon (\sigma+\ell) P^\nu \left( F_{x x} \right)_{\nu \mu}- \epsilon (\sigma+\ell) \Gamma^\alpha_{\beta \nu} P_\alpha P^\beta \left( F_{x p} \right)\indices{^\nu_\mu}. \end{split}
\end{align}

\section*{References}
\bibliography{reference2.bib} 
\end{document}